\documentclass{quantumview}
\usepackage[utf8]{inputenc}
\usepackage[english]{babel}
\usepackage[T1]{fontenc}
\usepackage{amsmath}
\usepackage{hyperref}
\usepackage{fixltx2e}
\usepackage[numbers,sort&compress]{natbib}

\begin{document}
	
	\title{Solve single photon detector problems}
	\author{Hao Shu}
	\email{$Hao\_B\_Shu@163.com$}
	\affiliation{Shenzhen University\\South China University of Technology}
	\date{}
	\orcid{0000-0002-9332-7103}
	
	\maketitle
	
	\begin{abstract}
	Single photon detector(SPD) problems arise in most quantum tasks, especially for measuring states going through high-lost channels. They are particularly prominent in quantum key distribution(QKD), which could be the most significant application in quantum information theory. In recent years, QKD distance has been improved dramatically but is still restricted because the bit error rate(QBER) caused by SPD dark counts will be out of control as the distance increases. If this problem can be solved, QKD can be implemented over arbitrarily long distances. However, previous solutions often result in impractical requirements such as superconductors while they can only reduce the dark count rate to finite low levels. In this paper, we solve SPD problems with today's technologies only. Although it is the no-cloning theorem that prevents a state from being measured multiple times to obtain a more reliable result, we propose a scheme circumventing the no-cloning theorem in certain tasks to allow a single state to be employed several times. The scheme demonstrates that imperfect detectors can provide nearly perfect results, namely, the QBER caused by dark counts can be reduced to arbitrarily low while in the meantime, detective efficiency can be improved to arbitrarily high. Consequently, QKD distance is not limited by the imperfect SPD anymore and can be improved from hundreds of kilometers to thousands without high-technology detectors. Furthermore, similar schemes can be applied for reducing measurement errors or improving the performance of sources. Finally, it is worth noting that although the paper is mainly discussed in the context of QKD, our scheme is an independent scheme that could be employed in other protocols wherever SPD are employed.
		\\
		\par\textbf{Keywords:} Long distance Quantum key distribution; Dark count; C-NOT gate; Single photon detector; Detective efficiency.
	\end{abstract}
	
	\maketitle

\section{Introduction}

	Quantum key distribution, firstly published in 1984\cite{BB1984Quantum} with substantial developments in the following dacades\cite{E1991Quantum,B1992Quantum,BB1992Quantum,GV1995Quantum,LY2018Overcoming,SP2000Simple,MZ2018Phase,LC2012Measurement,S2021Quantum,S2021Asymptotically,GL2004Security}, could be the most significant application in quantum information theory. Nowadays, the main problem is implementing practically, in which imperfect devices lead to errors. Among these, single photon detector(SPD) problems are considered as a main obstacle, which has not been solved yet. In principle, the QKD distance is bounded by the bit error rate(QBER), which is mainly determined by $\frac{t}{p}$, where $t=10^{-\frac{\alpha l}{10}}$ is the transmission rate depending on the distance $l$ and the loss coefficient $\alpha$ while $p$ is the dark count rate depending on the SPD\footnote{For example, in BB84 protocol, the QBER caused by dark counts approximates to $\frac{(1-t)p}{t\eta+2(1-t)p}\approx\frac{1}{\eta\frac{t}{p}+2}$, where $t,p\ll 1$ in long-distance communications.}. After a long-distance transmission in a lossy channel, the QBER approximates to $50\%$ since $\frac{t}{p}$ is close to 0, in which the security threshold is not satisfied. Theoretically speaking, QKD distance will not be limited if the dark count problem can be solved. On the other hand, another less important but still meaningful problem is the detective efficiency of SPD. It is often low, especially for low dark count ones.
	
	For 1550nm wavelength, $\alpha=0.2dB/km$, dark count rate of an InGaAs/InP SPD can be $1.36\times10^{-6}$ with $27.5\%$ detective efficiency at 223K operation temperature\cite{JL20171.25} while dark count rate of an upconversion SPD can be $4.6\times10^{-4}$ with $59\%$ detective efficiency at 300K operation temperature\cite{AW2004Efficient}, which imply that the distance of BB84 protocol\cite{BB1984Quantum} is at most nearly 220km and 110km, respectively\footnote{The numbers here are accurate to 10km while more exactly, the distances are about 222.75km and 112.87km instead of 220km and 110km, respectively. Here, we assume that other operations are perfect and take 11\% as the maximal error rate that can be tolerated.}. To improve the distance, a direct approach is reducing the dark count rate. However, SPD with high efficiency and low dark count rate often require superconductors which result in impractical operation temperatures\cite{Y2020Superconducting,WM2009Superconducting,ZJ2019Saturating} while even so, the dark count problem can only be reduced to a finite low-level but not be solved.
	
	Can the SPD problem be solved without replacing them? The answer is positive. Here we present a scheme that employs a copy strategy with commercial SPD only, solving the dark count problem as well as the detective efficiency problem. Our scheme demonstrates that imperfect SPD can provide nearly perfect results, namely QBER caused by dark counts (in fact, as well as due to imperfect measurements) can be reduced to arbitrarily low while detective efficiency can be improved to arbitrarily high. Furthermore, similar schemes can be applied for reducing measurement errors or removing empty signals caused by imperfect sources. Finally, it is worth noting that although the scheme is discussed in the context of QKD, it is in fact an independent scheme that could be employed in other protocols wherever SPD are employed.
	
	The key insights of the paper are that (1) Although the security of QKD schemes are promised by the no-cloning theorem, which in principle forbids copying an unknown state, the sifting procedure for each bit in QKD protocols always makes the state in exactly one chosen basis known after reconciliation effective while aborts it otherwise, which allows one circumvent the no-cloning theorem and copy the result states in some way. (2) If one can copy states in QKD protocols without influencing the security, a followed copy strategy might allow one to solve the SPD problems including the dark count problem as well as efficiency problems with any accuracy.
	
	\section{Scheme and performance}
	
	For simplicity, we assume that all devices are perfect except detectors unless especially pointed out. In particular, the legitimated partner (Alice and Bob) employs perfect single photon sources and the quantum channel is noiseless (but still encounters attenuations). Our scheme employs a copy strategy before detecting via SPD, see Figure 1. On one hand, it can remove empty signals since the outcome of empty ones is different from non-empty ones after copying. On the other hand, it can improve detective efficiency since there are more but coherent signals after copying. Here, the sources for copying can be coherent states with high intensity and do not need to be sent via channels, thus they encounter nearly no attenuations and are strong enough that the dark count effect on them can be ignored.
	
	\begin{figure*}
		\includegraphics[width=1.0\textwidth]{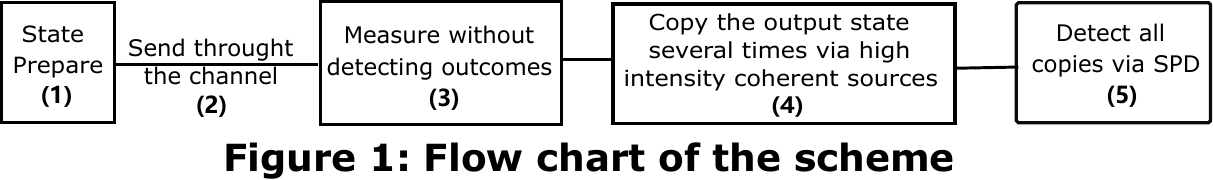}
	\end{figure*}

    We will employ operators $C_{0},C_{1},C_{+},C_{-}$, essentially C-NOT gates, as follow.

\begin{footnotesize}
$\begin{matrix}
	C_{0_{xy}}: &C^{2}\otimes C^{2} \rightarrow  C^{2}\otimes C^{2}\\
	&|0\rangle_{x}|0\rangle_{y} \rightarrow|0\rangle_{x}|0\rangle_{y}\\ 	
	&|1\rangle_{x}|0\rangle_{y} \rightarrow|1\rangle_{x}|1\rangle_{y}\\
	&|0\rangle_{x}|1\rangle_{y} \rightarrow|0\rangle_{x}|1\rangle_{y}\\ 	
	&|1\rangle_{x}|1\rangle_{y} \rightarrow|1\rangle_{x}|0\rangle_{y}
\end{matrix}$\qquad\qquad\quad
$\begin{matrix}
	C_{1_{xy}}: &C^{2}\otimes C^{2} \rightarrow  C^{2}\otimes C^{2}\\
	&|0\rangle_{x}|1\rangle_{y} \rightarrow  |0\rangle_{x}|0\rangle_{y}\\
	&|1\rangle_{x}|1\rangle_{y} \rightarrow  |1\rangle_{x}|1\rangle_{y}\\
	&|0\rangle_{x}|0\rangle_{y} \rightarrow|0\rangle_{x}|1\rangle_{y}\\ 	
	&|1\rangle_{x}|0\rangle_{y} \rightarrow|1\rangle_{x}|0\rangle_{y}
\end{matrix}$

$\begin{matrix}
	C_{+_{xy}}:&C^{2}\otimes C^{2} \rightarrow  C^{2}\otimes C^{2}\\
	&|+\rangle_{x}|+\rangle_{y} \rightarrow  |+\rangle_{x}|+\rangle_{y}\\ 	
	&|-\rangle_{x}|+\rangle_{y} \rightarrow  |-\rangle_{x}|-\rangle_{y}\\
	&|+\rangle_{x}|-\rangle_{y} \rightarrow|+\rangle_{x}|-\rangle_{y}\\ 	
	&|-\rangle_{x}|-\rangle_{y} \rightarrow|-\rangle_{x}|+\rangle_{y}
\end{matrix}$\qquad\qquad
$\begin{matrix}
	C_{-_{xy}}:&C^{2}\otimes C^{2} \rightarrow  C^{2}\otimes C^{2}\\
	&|+\rangle_{x}|-\rangle_{y} \rightarrow  |+\rangle_{x}|+\rangle_{y}\\
	&|-\rangle_{x}|-\rangle_{y} \rightarrow  |-\rangle_{x}|-\rangle_{y}\\
	&|+\rangle_{x}|+\rangle_{y} \rightarrow|+\rangle_{x}|-\rangle_{y}\\ 	
	&|-\rangle_{x}|+\rangle_{y} \rightarrow|-\rangle_{x}|+\rangle_{y}
\end{matrix}$
\end{footnotesize}

\noindent where $|0\rangle,|1\rangle,|+\rangle,|-\rangle\in C^{2}$ are states employed in BB84 protocol, and $x,y$ denote the partite.

\subsection{For BB84-like protocols}

Let us first take the BB84 protocol as an example. Our scheme is described as follows. The legitimated partner implements steps before measurement in BB84 protocol in the first place, in which Alice sends encoded states to Bob. When Bob receives a pulse, he chooses a basis to measure and copies the output state via C-NOT gates and sources with orthogonal polarization $d$ times, obtaining $d+1$ states, before detecting. Then he detects all $d+1$ states via SPD after measuring their polarization again and determines the bit by $m$ expected outcomes.

Precisely, for example, if Bob chooses basis $\{|0\rangle,|1\rangle\}$ for measuring (the first PBS, in the up-side in Figure 2) and assume that the output state in the up-side is $|0\rangle_{A}$ while the output state in right-side is $|1\rangle_{A}$, then he employs sources with polarization $|1\rangle_{S}$ in up-side and copies the output state via C-NOT gates $C_{1_{AS}}$, while he employs sources $|0\rangle_{S}$ in the right-side and copies the output state via C-NOT gates $C_{0_{AS}}$, where $A, S$ denote the partite of ordinary state and copying source respectively. He considers the bit is 0 if obtains at least $m$ outcome 0 in the up-up-side (after the second PBS, in the up-up-side) but obtains at most $m-1$ outcome 1 in the up-right-side (after the second PBS, in the up-right-side), while he considers the bit is 1 if he obtains at least $m$ outcome 1 in the up-right-side (after the second PBS, in the up-right-side) but obtains at most $m-1$ outcome 0 in the up-up-side (after the second PBS, in the up-up-side). Other steps are the same as in the BB84 protocol. See Figure 2.

\begin{figure}[h]
	\includegraphics[width=1\textwidth]{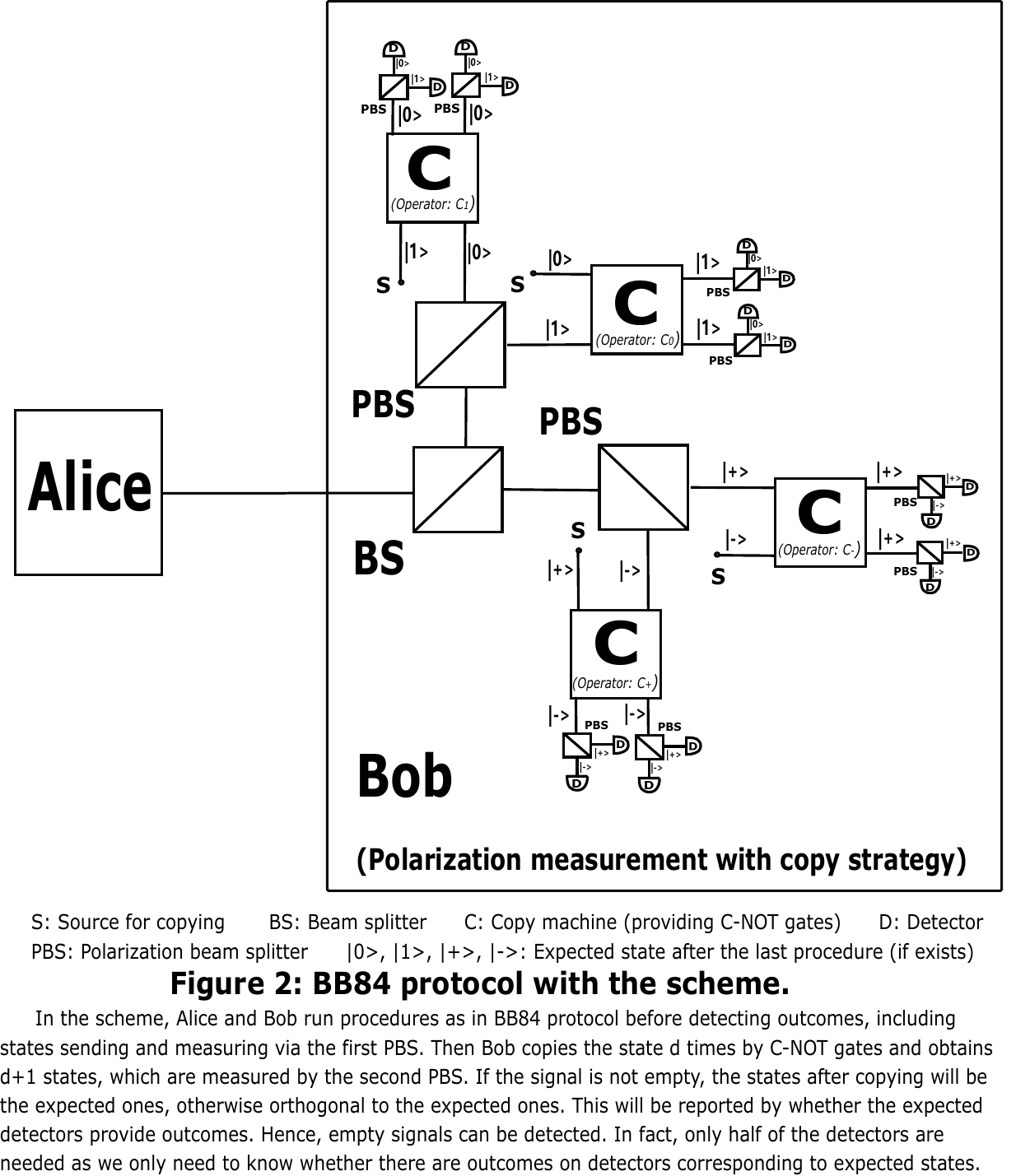}
\end{figure}

The QBER caused by dark counts can be estimated as
\begin{equation}
	Q_{det}\approx\frac{1}{\frac{t}{p^{m}}\frac{\sum_{k=m}^{d+1}\binom{d+1}{k}\eta^{k}(1-\eta)^{d+1-k}}{\binom{d+1}{m}}+2}
\end{equation}
 which increases as $\frac{t}{p^{m}}$ decreases, instead of $\frac{t}{p}$, providing $\binom{d+1}{i}p\ll 1, t\ll 1$ for all $i$ (for example $d\leq 8$), where $\eta$ is single SPD detective efficiency. By choosing $m$ larger, the QBER can be arbitrarily low at any distance. Details will be provided in supplied materials.
 
 The same idea can directly apply to other BB84-like protocols such as coding in both phase and polarization (Let us call it $'Phase-polarization\ protocol'$, see Figure 3).

\begin{figure}[]
	\includegraphics[width=1\textwidth]{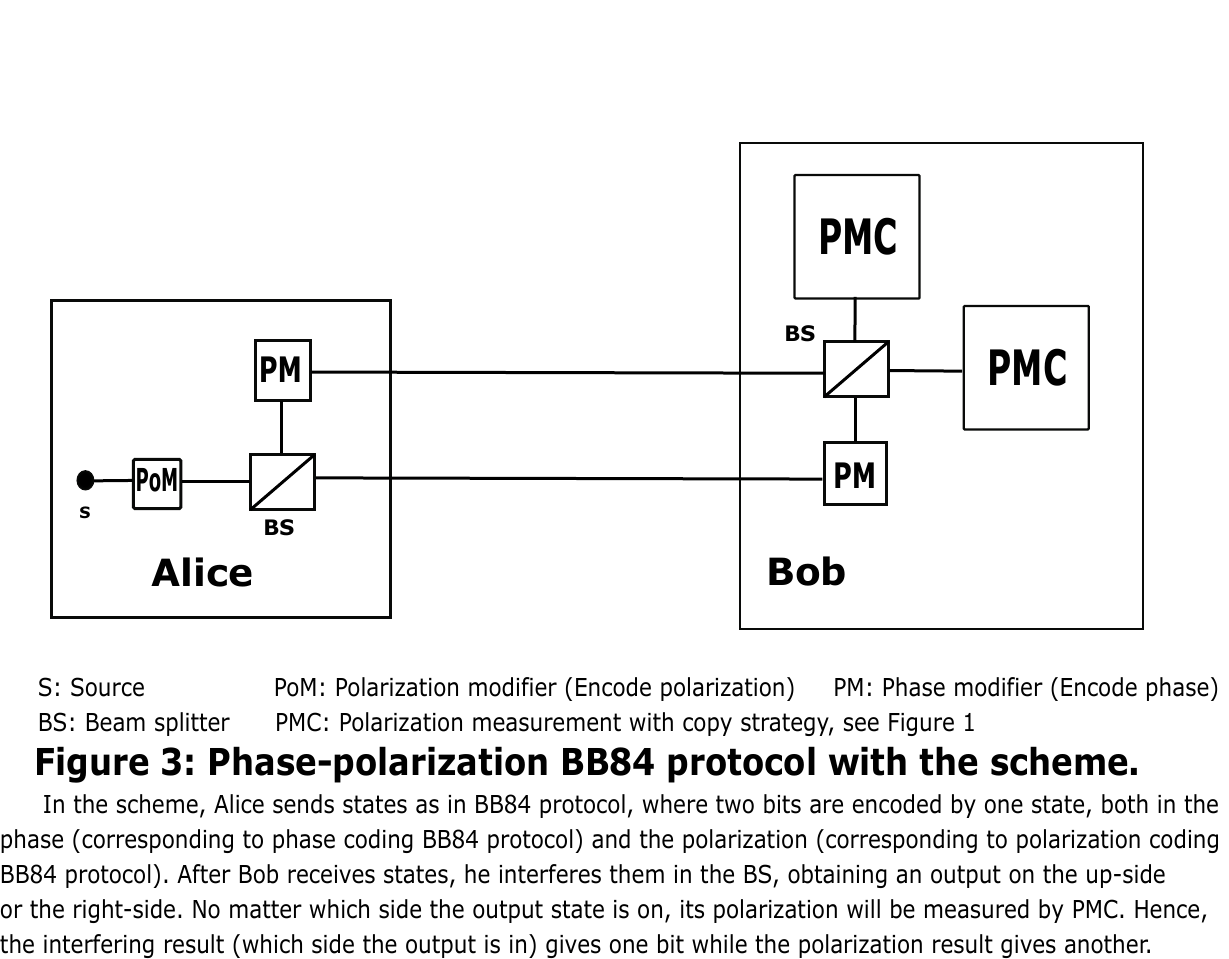}
	\\
	
	\includegraphics[width=1.0\textwidth]{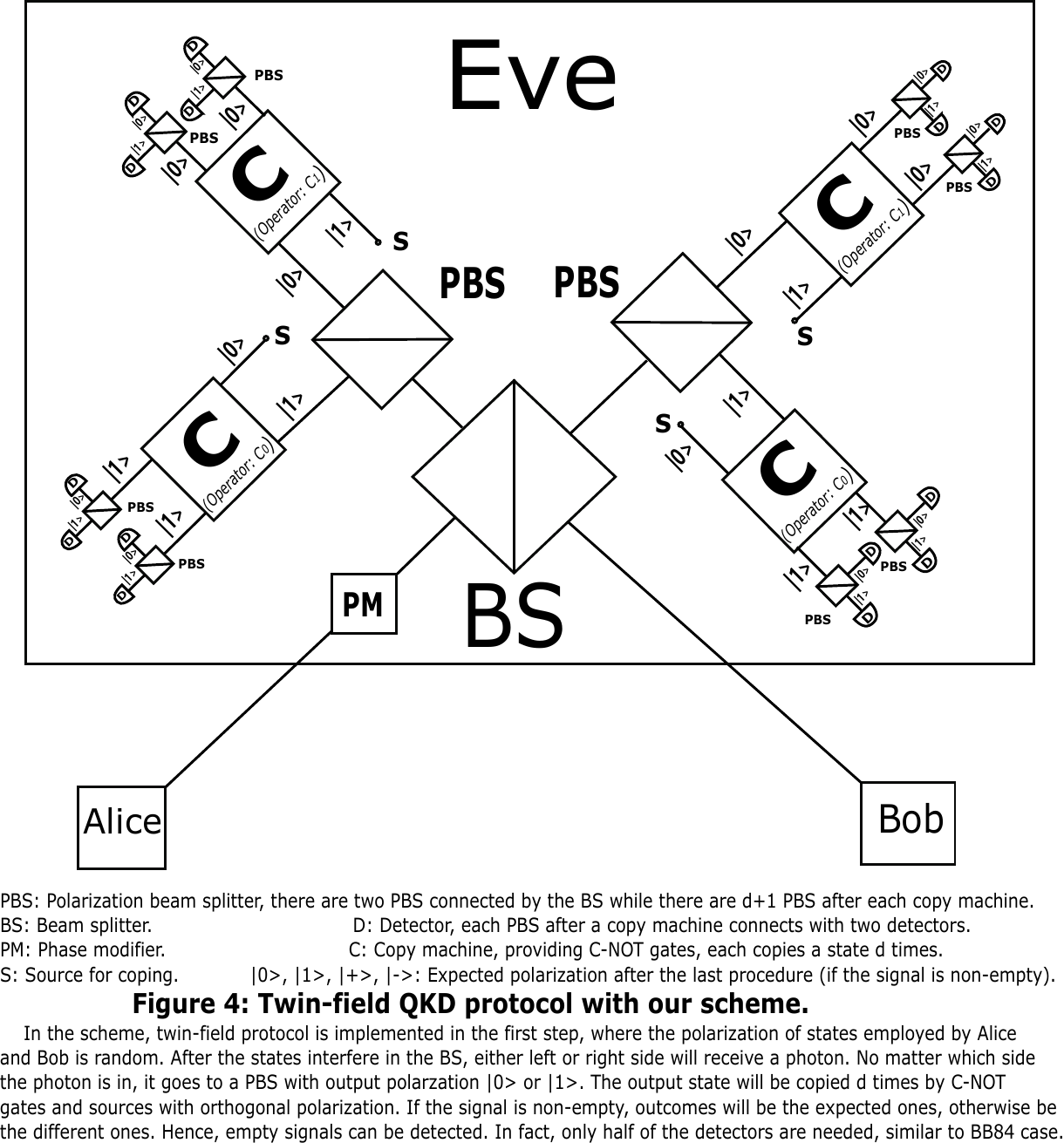}
\end{figure}

\subsection{For Twin-field protocol}

Although the scheme improves QKD distance for BB84 protocol, the key rate is still bounded by the rate-distance bounds\cite{PG2009Direct,TG2014Fundamental,PL2017Fundamental}. However, it could be modified for protocols that can break the bounds\cite{LY2018Overcoming,MZ2018Phase}. For applying the scheme to twin-field(TF) QKD protocol\cite{LY2018Overcoming}, see Figure 4.

In detail, let $|0\rangle, |1\rangle$ represent two orthogonal polarizations. Firstly, Alice and Bob implement TF protocol, sending states to untrusted third party Eve. For Eve, after interfering them in the BS, she sends the signal to a PBS and copies the output after PBS $d$ times via C-NOT gates $C_{0_{AE}}$ or $C_{1_{AE}}$ and sources with orthogonal polarization, where $A, E$ denote the partite of ordinary state and state for copying respectively. Then she measures all copies via polarization. If the signal is non-empty, expected polarization outcomes will be reported (marked in Figure 4), otherwise, the outcomes on auxiliary partite will be different from the expected ones. Only bits with at least $m$ expected outcomes on one side, namely, bits reporting at least $m$ $|0\rangle$ in detectors on one of the up-left-side and up-right-side, or bits reporting at least $m$ $|1\rangle$ in detectors on one of the down-left-side and down-right-side, with any other side reports at most $m-1$ expected outcomes, are considered as effective. Note that the final outcome of Eve is still given by the interfering result, namely the output state is on the left-side or right-side after BS, but not polarization results.

Hence, the QBER caused by dark counts is estimated as
 \begin{equation}
 	Q_{det}\approx\frac{1}{\frac{\sqrt{t}}{p^{m}}\frac{\sum_{k=m}^{d+1}\binom{d+1}{k}\eta^{k}(1-\eta)^{d+1-k}}{\binom{d+1}{m}}+2}
 \end{equation}
which increases as $\frac{\sqrt{t}}{p^{m}}$ decreases, providing $\binom{d+1}{i}p\ll 1, t\ll 1$ for all $i$, where $\eta$ is single SPD detective efficiency. Details will be provided in supplied materials.

\subsection{Performance}

In Figure 5 and Figure 6, the performances of the protocols with different parameters are presented. By employing nine copies ($d=8$) with effective events be those with at least $m$ expected outcomes, our scheme improves the distance of BB84 protocol from 220km to 450km, 720km, and over 800km for InGaAs/InP SPD (223K), while from 110km to 210km, 350km, and over 600km for Upconversion SPD (300K), respectively corresponding to $m=2,3$, and 5. As for TF protocol, the distance is improved from about 440km to about 890km, 1420km and over 2500km for InGaAs/InP SPD (223K), while from about 220km to about 420km, 720km, and over 1200km for upconversion SPD (300K), respectively corresponding to $m=2,3$, and 5\footnote{The numbers here are accurate to 10km}. Moreover, the key rate of the protocols can be higher than the ordinary ones, even in short distances.

\begin{figure}[]
	\includegraphics[width=1.0\textwidth]{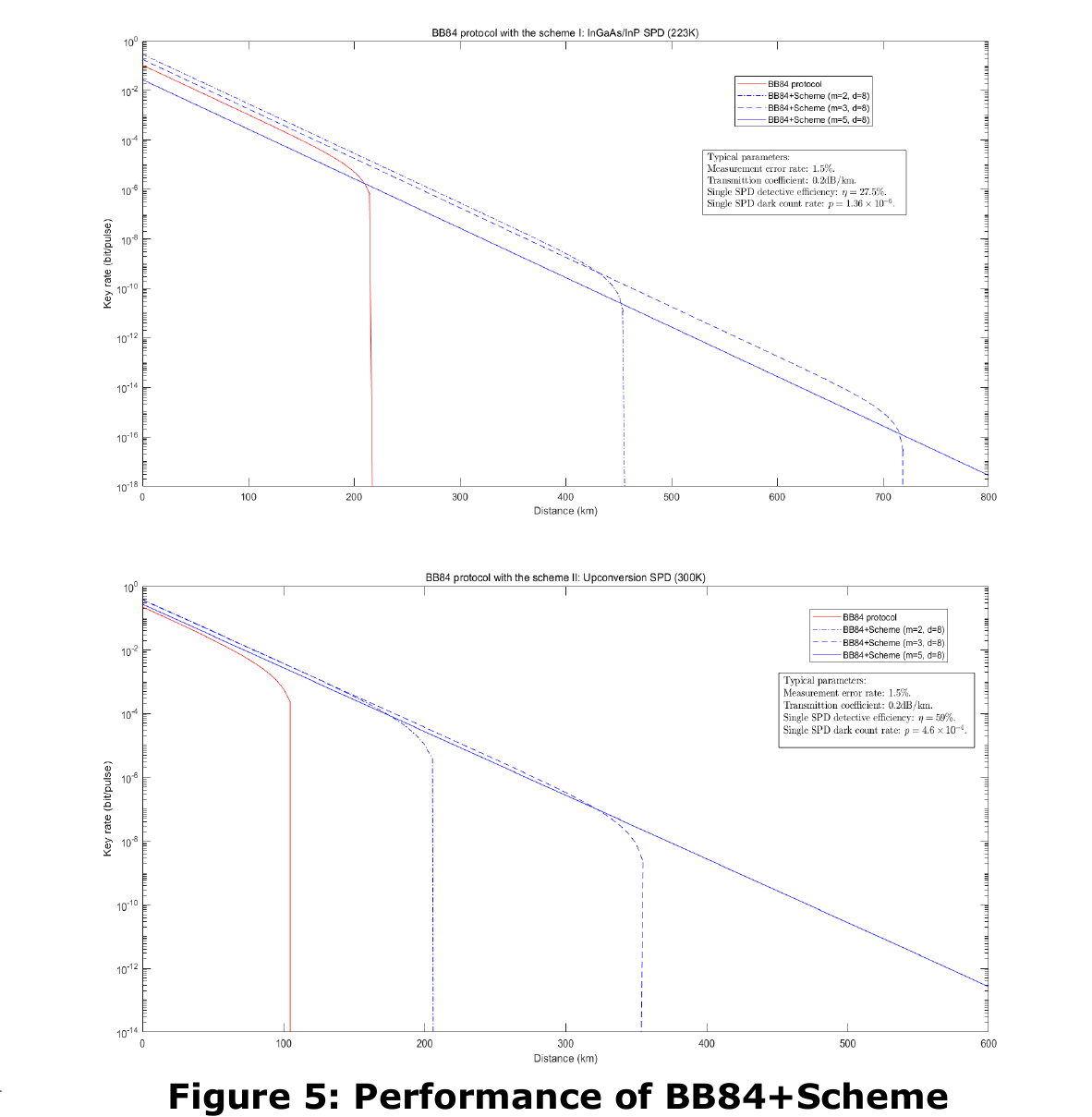}
\end{figure}

\begin{figure}[]
	\includegraphics[width=1.0\textwidth]{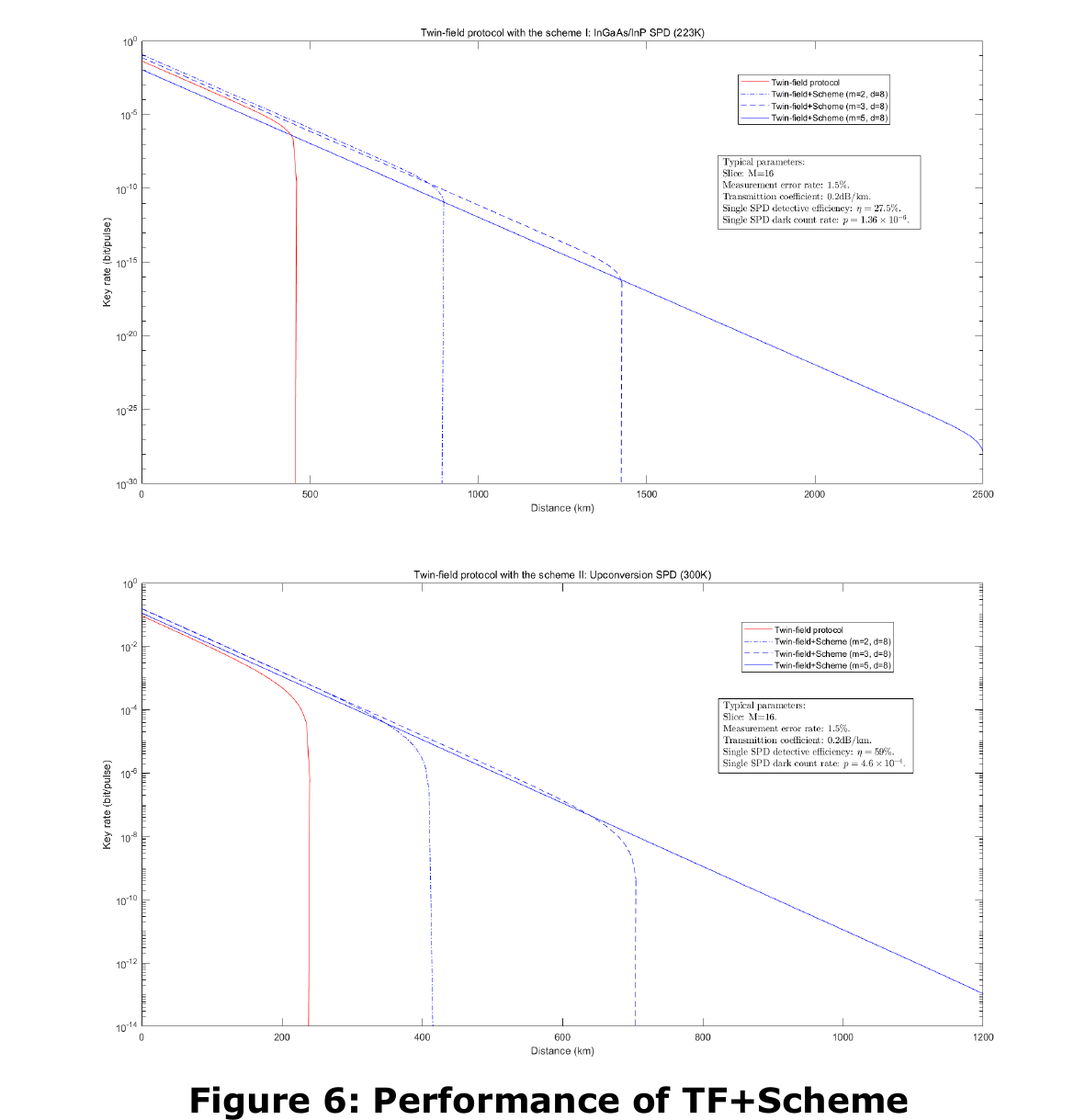}
\end{figure}

\section{Discussion}

\subsection{Improve detective efficiency}

For $d,m,\eta$ above, the whole detective efficiency will be
\begin{equation}
	1-\sum_{k=d+2-m}^{d+1}\binom{d+1}{k}(1-\eta)^{k}\eta^{d+1-k}
\end{equation}
For example, let $d=8, m=5$ and single SPD efficiency be $\eta=59\%$, then the whole detective efficiency exceeds $71\%$. By enlarging $d$, the whole detective efficiency can arbitrarily approximate 1.

\subsection{Reduce measurement error rate}

Furthermore, a similar scheme can reduce the whole measurement error rate. As an example, for the BB84 protocol, let Bob copy the state before measuring in the same basis chosen for measuring, and then measure all copies, see Figure 7.

\begin{figure*}
	\includegraphics[width=1.0\textwidth]{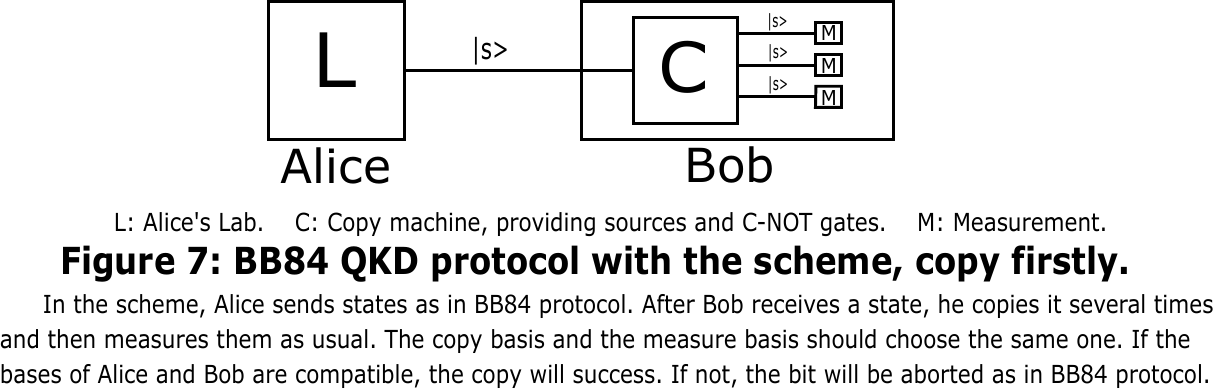}
\end{figure*}

 Precisely, in qubit case, for each time Bob copies the state, he employs two auxiliary partite $B_{1},B_{2}$ and copies the state twice. If he chooses to measure via basis $\{|0\rangle, |1\rangle\}$, he employs $|0\rangle_{B_{1}},|1\rangle_{B_{2}}$ as auxiliary states and operates $C_{0_{AB_{1}}}$ on partite $A,B_{1}$, $C_{0_{AB_{2}}}$ on partite $A,B_{2}$, respectively. If he chooses to measure via basis $\{|+\rangle, |-\rangle\}$, he employs $|+\rangle_{B_{1}},|-\rangle_{B_{2}}$ as auxiliary states and operates $C_{+_{AB_{1}}}$ on $A,B_{1}$ and $C_{-_{AB_{2}}}$ on $A,B_{2}$, respectively. Since in the sifting procedure, bits with incompatible basis will be discarded, without loss generality, assume that Alice sends state $|s\rangle \in \{|0\rangle,|1\rangle\}$ while Bob chooses basis $\{|0\rangle,|1\rangle\}$ for measuring. Hence, after copying, Bob's state would be $|s\rangle_{A}|s\rangle_{B_{1}}|s\rangle_{B_{2}}|s\rangle_{B_{1}'}|s\rangle_{B_{2}'}...$ if he receives the state successfully while the state would be $|\emptyset\rangle_{A}|0\rangle_{B_{1}}|1\rangle_{B_{2}}|0\rangle_{B_{1}'}|1\rangle_{B_{2}'}...$ if it was lost in the transmission, where $|\emptyset\rangle$ represents the empty signal. Let Bob copy the state $d_{0}$ times, obtaining $2d_{0}+1$ states which are divided into $d_{0}+1$ groups such that states of $B_{1},B_{2}$ in each copy are in the same group while the ordinary one is itself be a group. Hence, he provides $2d_{0}+1$ measurements via basis $\{|0\rangle, |1\rangle\}$. Groups with $B_{1}$ reports an outcome $|1\rangle$ (it is considered a group reporting 1) or $B_{2}$ reports an outcome $|0\rangle$ (it is considered as a group reporting 0) are effective while the group consists of the ordinary part is effective if it reports an outcome $|0\rangle$ or $|1\rangle$ (it is considered as a group reporting its outcome). Bob only keeps bits satisfying one of the following conditions:
 
 (1) There exist $m_{0}$ effective groups reporting outcome 0 while no $m_{0}$ groups reporting outcome 1 exist. In such a case, Bob considers the bit is 0.
 
 (2) There exist $m_{0}$ effective groups reporting outcome 1 while no $m_{0}$ groups reporting outcome 0 exist. In such a case, Bob considers the bit is 1.
 
 The qutrit case, namely Alice and Bob employ quantum system $C^{3}$ with an orthonormal basis $\{|0\rangle,|1\rangle,|2\rangle\}$, is easier to describe. In such case, Alice and Bob encode via basis $\{|0\rangle, |1\rangle\}$ or $\{|+\rangle, |-\rangle\}$ in a 2-dimensional subspace. However, Bob employs $|2\rangle$ as auxiliary states in copy procedure and measures via basis $\{|0\rangle, |1\rangle, |2\rangle \}$ or $\{|+\rangle, |-\rangle, |2\rangle\}$. In detail, we will employ the operators as follows.

\begin{footnotesize}
     $\begin{matrix}
		U_{xy}: &C^{3}\otimes C^{3} \rightarrow  C^{3}\otimes C^{3}\\
		&|0\rangle_{x}|2\rangle_{y} \rightarrow  |0\rangle_{x}|0\rangle_{y}\\ 	
		&|1\rangle_{x}|2\rangle_{y} \rightarrow  |1\rangle_{x}|1\rangle_{y}\\
		&|2\rangle_{x}|2\rangle_{y} \rightarrow  |2\rangle_{x}|2\rangle_{y}\\ 	
		&|0\rangle_{x}|1\rangle_{y} \rightarrow  |0\rangle_{x}|2\rangle_{y}\\
		&|1\rangle_{x}|1\rangle_{y} \rightarrow  |1\rangle_{x}|0\rangle_{y}\\ 	
		&|2\rangle_{x}|1\rangle_{y} \rightarrow  |2\rangle_{x}|1\rangle_{y}\\
		&|0\rangle_{x}|0\rangle_{y} \rightarrow  |0\rangle_{x}|1\rangle_{y}\\ 	
		&|1\rangle_{x}|0\rangle_{y} \rightarrow  |1\rangle_{x}|2\rangle_{y}\\
		&|2\rangle_{x}|0\rangle_{y} \rightarrow  |2\rangle_{x}|0\rangle_{y}
	\end{matrix}$\qquad\qquad\quad
	$\begin{matrix}
		V_{xy}: &C^{3}\otimes C^{3} \rightarrow  C^{3}\otimes C^{3}\\
		&|+\rangle_{x}|2\rangle_{y} \rightarrow  |+\rangle_{x}|+\rangle_{y}\\ 	
		&|-\rangle_{x}|2\rangle_{y} \rightarrow  |-\rangle_{x}|-\rangle_{y}\\
	    &|2\rangle_{x}|2\rangle_{y} \rightarrow  |2\rangle_{x}|2\rangle_{y}\\ 	
        &|+\rangle_{x}|+\rangle_{y} \rightarrow  |+\rangle_{x}|-\rangle_{y}\\
        &|-\rangle_{x}|+\rangle_{y} \rightarrow  |-\rangle_{x}|2\rangle_{y}\\ 	
        &|2\rangle_{x}|+\rangle_{y} \rightarrow  |2\rangle_{x}|+\rangle_{y}\\
        &|+\rangle_{x}|-\rangle_{y} \rightarrow  |+\rangle_{x}|2\rangle_{y}\\ 	
        &|-\rangle_{x}|-\rangle_{y} \rightarrow  |-\rangle_{x}|+\rangle_{y}\\
        &|2\rangle_{x}|-\rangle_{y} \rightarrow  |2\rangle_{x}|-\rangle_{y}
	\end{matrix}$
\end{footnotesize}

Assume that Bob employs $d$ auxiliary partite while Alice sends $|s\rangle_{A}$. For each auxiliary part $B$, Bob operates $U_{AB}$ if he chooses basis $\{|0\rangle, |1\rangle, |2\rangle \}$ or $V_{AB}$ if he chooses basis $\{|+\rangle, |-\rangle, |2\rangle \}$. Here, the $d+1$ copies including the ordinary one are divided into exactly $d+1$ groups while a measurement reporting one and only one outcome in $\{|0\rangle, |1\rangle\}$ or $\{|+\rangle, |-\rangle\}$ is effective. Without loss generality, let Alice send $|s\rangle_{A}\in\{|0\rangle, |1\rangle\}$ and Bob chooses $\{|0\rangle, |1\rangle\, |2\rangle\}$ for measuring. Hence, Bob only keeps bits satisfying one of the following conditions:

(1) There exist $m$ effective groups reporting outcome 0 while no $m$ groups reporting outcome 1 exist. In such a case, Bob considers the bit is 0.

(2) There exist $m$ effective groups reporting outcome 1 while no $m$ groups reporting outcome 0 exist. In such a case, Bob considers the bit is 1.

In qubit case, the QBER caused by dark counts is
\begin{equation}
	Q_{det}\approx\frac{1}{\frac{t}{p^{m_{0}}}\frac{\binom{d_{0}}{m_{0}-1}\eta^{m_{0}}(1-\eta)^{d_{0}-m_{0}+1}+\sum_{k=m_{0}}^{d_{0}}\binom{d_{0}}{k}\eta^{k}(1-\eta)^{d_{0}-k}}{\binom{d_{0}}{m_{0}-1}+\binom{d_{0}}{m_{0}}}+2}
\end{equation}
which increases as $\frac{t}{p^{m_{0}}}$ decreases, while in qutrit case, it is
\begin{equation}
	Q_{det}\approx\frac{1}{\frac{t}{p^{m}}\frac{\sum_{k=m}^{d+1}\binom{d+1}{k}\eta^{k}(1-\eta)^{d+1-k}}{\binom{d+1}{m}}+2}
\end{equation}
which increases as $\frac{t}{p^{m}}$ decreases, providing $\binom{d_{0}}{i}p\ll 1$ and $\binom{d+1}{i}p\ll 1$ for all $i$.

Now, denote the single measurement error rate by $p_{opt}$. For $d, m, \eta$ as above, the whole measurement error rate will be
\begin{footnotesize}
\begin{equation}
	Q_{opt}=\frac{\sum_{k=max(m,d+2-m)}^{d+1}\binom{d+1}{k}p_{opt}^{k}(1-p_{opt})^{d+1-k}}{\sum_{k=max(m,d+2-m)}^{d+1}\binom{d+1}{k}p_{opt}^{k}(1-p_{opt})^{d+1-k}+\sum_{k=0}^{min(m-1,d+1-m)}\binom{d+1}{k}p_{opt}^{k}(1-p_{opt})^{d+1-k}}.
\end{equation}
\end{footnotesize}

By enlarging $d$ and taking suitable $m$, $Q_{opt}$ can be arbitrarily low. For example, let $d=8, m=5$ and $p_{opt}=1.5\%$, then the whole measurement error rate is below $10^{-7}$. The same arguments hold for $d_{0}, m_{0}$.

\subsection{Employ for source}

It is worth noting that the scheme can be modified, applying to sources for removing empty signals, too. For example, copy the source state by high-intensity coherent sources with orthogonal polarization $d$ times and measure the copying states (but not the ordinary one). If the copying states report $m$ outcomes different from the copying sources, the ordinary state will be employed for the task, otherwise, it will be aborted. Hence, the probability of detecting an empty signal is $1-p^{m}$ while the probability of aborting a non-empty signal approximates to
\begin{equation}
	\sum_{k=d-m+1}^{d}\binom{d}{k}(1-\eta)^{k}\eta^{d-k}
\end{equation}

For suitable $d,m$, most of the empty signals can be removed while most of the non-empty signals can be left, similar to the discussions above.

\subsection{Practical implement}

In practical implementation, the fidelity of C-NOT gates can exceed $99\%$\cite{KW2021Demonstration,NT2022Fast} while the error rate can be calculated together with the measurement error rate, just viewing them as measurement errors. The influence of their imperfections can be small enough to negligible for suitable $m$ and $d$, explained above. Though there might be a few distances from applying C-NOT gates in these protocols, one can imagine that it should not be a serious obstacle. Also note that even with low-efficiency C-NOT operations, the scheme might still be effective. To understand this, only notice that in the above tasks, we generally do not require all detectors to report the correct result but consider a measurement is successful if certain effective outcomes are gained. Therefore, depending on how much accuracy one needs, the scheme can tolerate some failures and even some errors of C-NOT gates.

For implementing these protocols by weak coherent sources, decoy state methods\cite{LM2005Decoy} can be applied directly, since our scheme has no affections on the transmission.

\subsection{Other notes}

The security of protocols is not affected by the scheme, since the scheme employs nothing but repeats the detective procedure. The scheme might also reduce problems caused by SPD dead time since it can tolerate that some measurements give no outcomes.

Note that while we choose a special source for each possible state in the copying procedure, the source might be chosen multiformly in practice to save states. For example, when employing $C_{0xy}$ for copying, the sources can also be $|1\rangle$ with a bit flip in every copying state.

Also note that the most significant insight in this paper is that while it is the no-cloning theorem that promises the security of QKD, states can be cloned in the measuring stage without affecting the security. The whole protocol can be viewed as standard QKD protocols assisted by cloning protocols. The employment of C-NOT gates for cloning is only because they might be the most understandable ones with substantial works and are enough in the abovementioned examples. Certainly, other cloning schemes might also be employed wherever needed, but it is beyond this paper.

\subsection{Apply in other QKD protocols and other tasks}

It is easy to see that the scheme can be extended to other QKD protocols such as DI-QKD(device-independent QKD) and MDI-QKD(measurement-device-independent QKD) as it is not so special for the examples presented above. Besides, the idea of the scheme can be employed in other tasks associating or not associating with QKD tasks. Here we specifically mention that it can be employed to remove a kind of Trojan horse attack. In \cite{S2022Measurement}, we have employed C-NOT gates to handle a kind of two-stage attack to protect the measurement-device-independenization of general MDI-QKD protocol.

In fact, as we have mentioned in the introduction, the presented scheme is an independent scheme although we employ several QKD tasks as examples. Indeed, all measurement tasks can somehow employ the scheme since, no matter what outcome a detector represents in the measurement, there exists an orthogonal state for it and thus it can be copied. Even if there are no losses in the channel, namely $t=1$, the scheme can also be employed for increasing detective efficiency, which should benefit nearly all measurement tasks. The same discussion also applies to the employment in sources, namely, it is not especially for QKD tasks but also for any task that should benefit by reducing empty signals.

\section{Conclusion}

In conclusion, we presented a scheme employing the copy strategy to solve SPD problems, which removes the distance restriction in implementing QKD tasks. With the scheme, nearly perfect results can be obtained by the imperfect SPD, namely, QBER caused by dark counts, measurements, and C-NOT gates can be reduced to arbitrary low, while detective efficiency can be increased to arbitrary high. The scheme is suitable for phase coding, polarization coding, or both, corresponding to the twin-field argument, BB84 argument, and phase-polarization argument above, respectively, while besides QKD, it might be generalized for other tasks such as removing empty signals in sources. With it, commercial SPD can be employed as perfect ones, which demonstrates that SPD might not be a problem anymore.

\bibliographystyle{quantum}
\bibliography{Bibliog}

\section{Supplied material}

Let us provide details of calculating the QBER for BB84 protocol and Twin-field protocol with the scheme. Here we assume that $t\ll 1$ and $\binom{d+1}{i}p\ll 1, \binom{d_{0}+1}{i}p\ll 1$ for all i (for example $d,d_{0}\leq8$) when calculating approximations. Note that the detectors are assumed to be independent of each other and thus the probabilities of them reporting an error outcome (caused by the dark count) are independent. As in the above sections, $p, \eta$ still denote the dark count rate and the measurement efficiency of a single SPD.

\subsection{BB84+scheme (measure firstly)}

Without loss generality, assume that Alice sends state $|0\rangle$ and Bob measures via basis $\{|0\rangle,|1\rangle\}$. The probabilities are calculated as follows.

Bob obtains a correct outcome on a single copy when the state is transmitted successfully (an up-up SPD reports outcome $|0\rangle$):
\\
$P_{t}'(0)=\eta+(1-\eta)p\approx\eta$;

Bob obtains an error outcome on a single copy when the state is transmitted successfully (an up-right SPD reports an outcome $|1\rangle$):
$P_{t}'(1)=p$;

Bob obtains a correct outcome on a single copy when the state is lost (an up-up SPD reports an outcome $|0\rangle$):
$P_{l}'(0)=p$;

Bob obtains an error outcome on a single copy when the state is lost (an up-right SPD reports an outcome $|1\rangle$):
$P_{l}'(1)=p$;

Bob obtains the correct outcome on a bit when the state is transmitted successfully (at least $m$ $|0\rangle$ in the up-up-side but at most $m-1$ $|1\rangle$ in the up-right-side):

\begin{footnotesize}
\noindent $P_{t}(0)=[\sum_{k=m}^{d+1}\binom{d+1}{k}P'_{t}(0)^{k}(1-P'_{t}(0))^{d+1-k}][1-\sum_{k=m}^{d+1}\binom{d+1}{k}P'_{t}(1)^{k}(1-P'_{t}(1))^{d+1-k}]
\\
\approx \sum_{k=m}^{d+1}\binom{d+1}{k}\eta^{k}(1-\eta)^{d+1-k}$;
\end{footnotesize}

Bob obtains the error outcome on a bit when the state is transmitted successfully (at least $m$ $|1\rangle$ in the up-right-side but at most $m-1$ $|0\rangle$ in the up-up-side):

\begin{footnotesize}
\noindent $P_{t}(1)=[\sum_{k=m}^{d+1}\binom{d+1}{k}P'_{t}(1)^{k}(1-P'_{t}(1))^{d+1-k}][1-\sum_{k=m}^{d+1}\binom{d+1}{k}P'_{t}(0)^{k}(1-P'_{t}(0))^{d+1-k}]
\\
\approx\binom{d+1}{m}p^{m}(1-\sum_{k=m}^{d+1}\binom{d+1}{k}\eta^{k}(1-\eta)^{d+1-k})$;
\end{footnotesize}

Bob obtains a correct outcome on a bit when the state is lost (at least $m$ $|0\rangle$ in the up-up-side but at most $m-1$ $|1\rangle$ in the up-right-side):
\\
\begin{footnotesize}
	$P_{l}(0)=[\sum_{k=m}^{d+1}\binom{d+1}{k}P'_{l}(0)^{k}(1-P'_{l}(0))^{d+1-k}][1-\sum_{k=m}^{d+1}\binom{d+1}{k}P'_{l}(1)^{k}(1-P'_{l}(1))^{d+1-k}]
	\\
	\approx\binom{d+1}{m}p^{m}$;
\end{footnotesize} 

Bob obtains an error outcome on a bit when the state is lost (at least $m$ $|1\rangle$ in the up-right-side but at most $m-1$ $|0\rangle$ in the up-up-side):
\\
\begin{footnotesize}
	$P_{l}(1)=[\sum_{k=m}^{d+1}\binom{d+1}{k}P'_{l}(1)^{k}(1-P'_{l}(1))^{d+1-k}][1-\sum_{k=m}^{d+1}\binom{d+1}{k}P'_{l}(0)^{k}(1-P'_{l}(0))^{d+1-k}]
	\\
	\approx\binom{d+1}{m}p^{m}$;
\end{footnotesize}

Successful event rate: $R_{sift}=\frac{1}{2}[t(P_{t}(0)+P_{t}(1))+(1-t)(P_{l}(0)+P_{l}(1))]$;

Successful but error event rate: $R_{det}=\frac{1}{2}[tP_{t}(1)+(1-t)P_{l}(1)]$;

Bit error rate caused by dark counts: 
\begin{footnotesize}
	$Q_{det}=\frac{R_{det}}{R_{sift}}\approx\frac{1}{\frac{t}{p^{m}}\frac{\sum_{k=m}^{d+1}\binom{d+1}{k}\eta^{k}(1-\eta)^{d+1-k}}{\binom{d+1}{m}}+2}$.
\end{footnotesize}

\subsection{TF+scheme}

Without loss generality, assume that the expected interfering output is on the left side and the output state after the first PBS (if exists) is $|0\rangle$.

Eve obtains an expected outcome on a left-up SPD ($|0\rangle$) when the state is transmitted successfully: $P'_{t}=\eta+(1-\eta)p\approx\eta$;

Eve obtains an expected outcome on any one of a left-down SPD ($|1\rangle$), a right-up SPD ($|0\rangle$),  and a right-down SPD ($|1\rangle$) when the state is transmitted successfully, or Bob obtains an expected outcome on any one SPD when the state is lost:
$P'_{l}=p$;

Eve obtains a final outcome on the left-up-side when the signal after interfering is non-empty (at least $m$ expected outcomes in left-up-side):
\\
$P_{t}=\sum_{k=m}^{d+1}\binom{d+1}{k}{P'_{t}}^{k}(1-P'_{t})^{d+1-k}\approx\sum_{k=m}^{d+1}\binom{d+1}{k}\eta^{k}(1-\eta)^{d+1-k}$;

Eve obtains a final outcome on any other side except the left-up-side when the signal after interfering is non-empty or obtains a final outcome on any side when the signal after interfering is empty (at least $m$ expected outcomes in the corresponding side):
$P_{l}=\sum_{k=m}^{d+1}\binom{d+1}{k}{P'_{l}}^{k}(1-P'_{l})^{d+1-k}\approx\binom{d+1}{m}p^{m}$;

Eve obtains the correct outcome on a bit when the state is transmitted successfully (Eve obtains a final outcome on one of the left-up and left-down sides but not on any one of the right-up and right-down sides):
\\
$P_{t}(correct)=[(1-P_{t})P_{l}+P_{t}(1-P_{l})](1-P_{l})^{2}\approx P_{t}$;

Eve obtains the error outcome on a bit when the state is transmitted successfully (Eve obtains a final outcome on one of the right-up, and right-down sides but not on any one of the left-up and left-down sides):
\\
$P_{t}(error)=2(1-P_{t})(1-P_{l})(1-P_{l})P_{l}\approx 2(1-P_{t})P_{l}$;

Eve obtains an outcome on a bit when the state is lost (Eve obtains a final outcome on one and only one of the left-up, left-down, right-up, and right-down sides):
$P_{l}(correct)=P_{l}(error)=2(1-P_{l})^{3}P_{l}\approx 2P_{l}$;

Transmission rate: $t'=1-(1-\sqrt{t})^{2}=2\sqrt{t}-t\approx 2\sqrt{t}$;

Successful event rate (M=16):
\\ $R_{sift}=\frac{2}{16}[t'(P_{t}(correct)+P_{t}(error))+(1-t')(P_{l}(correct)+P_{l}(error))]$;

Successful but error event rate: $R_{det}=\frac{2}{16}[t'P_{t}(error)+(1-t')P_{l}(error)]$;

Bit error rate caused by dark counts: $Q_{det}\approx\frac{1}{\frac{\sqrt{t}}{p^{m}}\frac{\sum_{k=m}^{d+1}\binom{d+1}{k}\eta^{k}(1-\eta)^{d+1-k}}{\binom{d+1}{m}}+2}$.

\subsection{BB84+scheme (copy before measure)}

\subsubsection{Qutrit case}

Without loss generality, assume that Alice sends state $|0\rangle$ while Bob chooses basis$\{|0\rangle, |1\rangle, |2\rangle\}$ for measuring.

Bob obtains outcome 0 on a single measurement when the state is transmitted successfully:

\noindent $P'_{t}(0)=[\eta+(1-\eta)p)](1-p)\approx\eta$;

Bob obtains outcome 1 on a single measurement when the state is transmitted successfully:

\noindent $P'_{t}(1)=(1-\eta)(1-p)p\approx(1-\eta)p$;

Bob obtains outcome 0 on a single measurement when the state is lost: $P'_{l}(0)=p(1-p)\approx p$;

Bob obtains outcome 1 on a single measurement when the state is lost: $P'_{l}(1)=p(1-p)\approx p$;

Bob obtains outcome 0 when the state is transmitted successfully:

\noindent
\begin{small}
	$P_{t}(0)=\sum_{k=m}^{d+1}\binom{d+1}{k}P'_{t}(0)^{k}\sum_{j=0}^{min(m-1,d+1-k)}\binom{d+1-k}{j}P'_{t}(1)^{j}(1-P'_{t}(0)-P'_{t}(1))^{d+1-k-j}
	\\
	\approx\sum_{k=m}^{d+1}\binom{d+1}{k}\eta^{k}(1-\eta)^{d+1-k}$;
\end{small} 

Bob obtains outcome 1 when the state is transmitted successfully:

\begin{footnotesize}
	\begin{equation*}
		\begin{aligned}
			P_{t}(1)&=\sum_{k=m}^{d+1}\binom{d+1}{k}P'_{t}(1)^{k}\sum_{j=0}^{min(m-1,d+1-k)}\binom{d+1-k}{j}P'_{t}(0)^{j}(1-P'_{t}(1)-P'_{t}(0))^{d+1-k-j}
			\\
			&\approx\binom{d+1}{m}[(1-\eta)p]^{m}\sum_{j=0}^{min(m-1,d+1-m)}\binom{d+1-m}{j}\eta^{j}(1-\eta)^{d+1-m-j};
		\end{aligned}
	\end{equation*}
\end{footnotesize}

Bob obtains outcome 0 when the state is lost:

\begin{small}
\noindent $P_{l}(0)=\sum_{k=m}^{d+1}\binom{d+1}{k}P'_{l}(0)^{k}\sum_{j=0}^{min(m-1,d+1-k)}\binom{d+1-k}{j}P'_{l}(1)^{j}(1-P'_{l}(0)-P'_{l}(1))^{d+1-k-j}
\\
\approx\binom{d+1}{m}p^{m}$;
\end{small}

Bob obtains outcome 1 when the state is lost:
$P_{l}(1)=P_{l}(0)$;

Successful event rate: $R_{sift}=\frac{1}{2}[t(P_{t}(0)+P_{t}(1))+(1-t)(P_{l}(0)+P_{l}(1))]$;

Successful but error event rate: $R_{det}=\frac{1}{2}[tP_{t}(1)+(1-t)P_{l}(1)]$;

Bit error rate caused by dark counts:
\begin{footnotesize}
 $Q_{det}=\frac{R_{det}}{R_{sift}}\approx\frac{1}{\frac{t}{p^{m}}\frac{\sum_{k=m}^{d+1}\binom{d+1}{k}\eta^{k}(1-\eta)^{d+1-k}}{\binom{d+1}{m}}+2}$.
\end{footnotesize}

\subsubsection{Qubit BB84 case}

Without loss generality, assume that Alice sends state $|0\rangle$ while Bob chooses basis$\{|0\rangle, |1\rangle\}$ for measuring.

Bob obtains outcome 0 on the ordinary part when the state is transmitted successfully:

\noindent $P_{t}''(0)=(\eta+(1-\eta)p)(1-p)\approx\eta$;

Bob obtains outcome 1 on the ordinary part when the state is transmitted successfully:

\noindent $P_{t}''(1)=(1-\eta)(1-p)p\approx(1-\eta)p$;

Bob obtains outcome 0 on a two-state group when the state is transmitted successfully:

\noindent $P_{t}'(0)=\eta+(1-\eta)p\approx\eta$;

Bob obtains outcome 1 on a two-state group when the state is transmitted successfully:
\\
$P_{t}'(1)=p$;

Bob obtains outcome 0 on the ordinary part when the state is lost:
$P_{l}''(0)=p$;

Bob obtains outcome 1 on the ordinary part when the state is lost:
$P_{l}''(1)=p$;

Bob obtains outcome 0 on a two-state group when the state is lost:
$P_{l}'(0)=p$;

Bob obtains outcome 1 on a two-state group when the state is lost:
$P_{l}'(1)=p$;

Bob obtains outcome 0 when the state is transmitted successfully:

\begin{footnotesize}
\begin{equation*}
	\begin{aligned}
		P_{t}(0)=&P_{t}''(0)\sum_{k=m_{0}-1}^{d_{0}}\binom{d_{0}}{k}P'_{t}(0)^{k}\sum_{j=0}^{min(m_{0}-1,d_{0}-k)}\binom{d_{0}-k}{j}P'_{t}(1)^{j}[(1-P'_{t}(0))(1-P'_{t}(1))]^{d_{0}-k-j}
		\\
		&+P_{t}''(1)\sum_{k=m_{0}}^{d_{0}}\binom{d_{0}}{k}P'_{t}(0)^{k}\sum_{j=0}^{min(m_{0}-2,d_{0}-k)}\binom{d_{0}-k}{j}P'_{t}(1)^{j}[(1-P'_{t}(0))(1-P'_{t}(1))]^{d_{0}-k-j}
		\\
		&+(1-P_{t}''(0))(1-P_{t}''(1))\sum_{k=m_{0}}^{d_{0}}\binom{d_{0}}{k}P'_{t}(0)^{k}\sum_{j=0}^{min(m_{0}-1,d_{0}-k)}\binom{d_{0}-k}{j}P'_{t}(1)^{j}[(1-P'_{t}(0))(1-P'_{t}(1))]^{d_{0}-k-j}
		\\
		\approx&\binom{d_{0}}{m_{0}-1}\eta^{m_{0}}(1-\eta)^{d_{0}+1-m_{0}}+\sum_{k=m_{0}}^{d_{0}}\binom{d_{0}}{k}\eta^{k}(1-\eta)^{d_{0}-k};
	\end{aligned}
\end{equation*}
\end{footnotesize}

Bob obtains outcome 1 when the state is transmitted successfully:

\begin{footnotesize}
\begin{equation*}
	\begin{aligned}
		P_{t}(1)=&P_{t}''(1)\sum_{k=m_{0}-1}^{d_{0}}\binom{d_{0}}{k}P'_{t}(1)^{k}\sum_{j=0}^{min(m_{0}-1,d_{0}-k)}\binom{d_{0}-k}{j}P'_{t}(0)^{j}[(1-P'_{t}(1))(1-P'_{t}(0))]^{d_{0}-k-j}
		\\
		&+P_{t}''(0)\sum_{k=m_{0}}^{d_{0}}\binom{d_{0}}{k}P'_{t}(1)^{k}\sum_{j=0}^{min(m_{0}-2,d_{0}-k)}\binom{d_{0}-k}{j}P'_{t}(0)^{j}[(1-P'_{t}(1))(1-P'_{t}(0))]^{d_{0}-k-j}
		\\
		&+(1-P_{t}''(1))(1-P_{t}''(0))\sum_{k=m_{0}}^{d_{0}}\binom{d_{0}}{k}P'_{t}(1)^{k}\sum_{j=0}^{min(m_{0}-1,d_{0}-k)}\binom{d_{0}-k}{j}P'_{t}(0)^{j}[(1-P'_{t}(1))(1-P'_{t}(0))]^{d_{0}-k-j}
		\\
		\approx&(1-\eta)p^{m_{0}}\binom{d_{0}}{m_{0}-1}\sum_{j=0}^{min(m_{0}-1,d_{0}-m_{0}+1)}\binom{d_{0}-m_{0}+1}{j}\eta^{j}(1-\eta)^{d_{0}-m_{0}+1-j}
		\\
		&+\eta\binom{d_{0}}{m_{0}}p^{m_{0}}\sum_{j=0}^{min(m_{0}-2,d_{0}-m_{0})}\binom{d_{0}-m_{0}}{j}\eta^{j}(1-\eta)^{d_{0}-m_{0}-j}
		\\
		&+(1-\eta)\binom{d_{0}}{m_{0}}p^{m_{0}}\sum_{j=0}^{min(m_{0}-1,d_{0}-m_{0})}\binom{d_{0}-m_{0}}{j}\eta^{j}(1-\eta)^{d_{0}-m_{0}-j};
	\end{aligned}
\end{equation*}
\end{footnotesize}

Bob obtains outcome 0 when the state is lost:

\begin{footnotesize}
\begin{equation*}
	\begin{aligned}
		P_{l}(0)=&P_{n}''(0)\sum_{k=m_{0}-1}^{d_{0}}\binom{d_{0}}{k}P'_{n}(0)^{k}\sum_{j=0}^{min(m_{0}-1,d_{0}-k)}\binom{d_{0}-k}{j}P'_{n}(1)^{j}[(1-P'_{n}(0))(1-P'_{n}(1))]^{d_{0}-k-j}
		\\
		&+P_{n}''(1)\sum_{k=m_{0}}^{d_{0}}\binom{d_{0}}{k}P'_{n}(0)^{k}\sum_{j=0}^{min(m_{0}-2,d_{0}-k)}\binom{d_{0}-k}{j}P'_{n}(1)^{j}[(1-P'_{n}(0))(1-P'_{n}(1))]^{d_{0}-k-j}
		\\
		&+(1-P_{n}''(0))(1-P_{n}''(1))\sum_{k=m_{0}}^{d_{0}}\binom{d_{0}}{k}P'_{n}(0)^{k}\sum_{j=0}^{min(m_{0}-1,d_{0}-k)}\binom{d_{0}-k}{j}P'_{n}(1)^{j}[(1-P'_{n}(0))(1-P'_{n}(1))]^{d_{0}-k-j}
		\\
		\approx&(\binom{d_{0}}{m_{0}-1}+\binom{d_{0}}{m_{0}})p^{m_{0}};
	\end{aligned}
\end{equation*}
\end{footnotesize}

Bob obtains outcome 1 when the state is lost: $P_{l}(1)=P_{l}(0)$;

Successful event rate: $R_{sift}=\frac{1}{2}[t(P_{t}(0)+P_{t}(1))+(1-t)(P_{l}(0)+P_{l}(1))]$;

Successful but error event rate: $R_{det}=\frac{1}{2}[tP_{t}(1)+(1-t)P_{l}(1)]$;

Bit error rate caused by dark counts:
\\
$Q_{det}=\frac{R_{det}}{R_{sift}}\approx\frac{1}{\frac{t}{p^{m_{0}}}\frac{\binom{d_{0}}{m_{0}-1}\eta^{m_{0}}(1-\eta)^{d_{0}-m_{0}+1}+\sum_{k=m_{0}}^{d_{0}}\binom{d_{0}}{k}\eta^{k}(1-\eta)^{d_{0}-k}}{\binom{d_{0}}{m_{0}-1}+\binom{d_{0}}{m_{0}}}+2}$.

\end{document}